\documentclass[twocolumn,showpacs,preprintnumbers,
amsmath,amssymb,aps,prd,nofootinbib,superscriptaddress,
eqsecnum]{revtex4}
\usepackage{graphicx}
\usepackage{dcolumn}
\makeatletter
\renewcommand{\p@subsection}{}
\makeatother

\newcommand{\Rmnum}[1]{\expandafter\@slowromancap\romannumeral #1@}

\newcommand{\be}{\begin{eqnarray}}
\newcommand{\ee}{\end{eqnarray}}

\def\lsim{\mathrel{\rlap{\lower3pt\hbox{\hskip1pt$\sim$}}
     \raise1pt\hbox{$<$}}} %less than or approx. symbol
\def\gsim{\mathrel{\rlap{\lower3pt\hbox{\hskip1pt$\sim$}}
     \raise1pt\hbox{$>$}}} %greater than or approx. symbol

\def\bi{\bibitem}

\begin{document}

\title{Hyperons and Condensed Kaons in Compact Stars}

\author{Hyun Kyu Lee}
\affiliation{%
Department of Physics, Hanyang University, Seoul 133-791, Korea}

\author{Mannque Rho}

\affiliation{%
Institut de Physique Th\'eorique,
CEA Saclay, 91191 Gif-sur-Yvette c\'edex, France \& \\
Department of Physics, Hanyang University, Seoul 133-791, Korea
}

\date{\today}

\begin{abstract}
Using the Callan-Klebanov bound state model for hyperons simulated on crystal lattice to describe strange baryonic matter, we argue that to ${\cal O} (1)$ in the large $N_c$ counting to which the theory is robust,  hyperons can figure only when -- or after -- kaons condense in compact-star matter. We also discuss how the skyrmion-half-skyrmion topological transition affects the equation of state (EoS) of dense baryonic matter. The observations made in this note open wide the issue of how to theoretically access the EoS of compact stars.
\end{abstract}

\maketitle

%%%%%%%%%%%%%%%%%%%%%%%%%%%%%%%%%%%%%%%%%%%%%%%%%%%%%

{\it Introduction:}
Strangeness can figure in compact stars in a very crucial way. The appearance of hyperons~\cite{hyperons} and/or condensed kaons~\cite{kaons} at a density above that of nuclear matter can soften the EoS for dense matter, influencing the maximum star mass and the cooling of the star formed in supernova explosion. There has been a long-standing debate as to what appears first in the gravitational compression of compact-star matter, hyperons or kaon condensation. If hyperons appear first, then they will substantially lower the electron chemical potential so that the electrons can no longer transform to kaons in the way discussed in \cite{BTKR}. This will then banish kaon condensation from the scenario. The alternative possibility -- which we think is more likely --  that kaon condensation can take place {\em before} the appearance of hyperons has not been explored up to date. It is the purpose of this paper to suggest that hyperons can appear not before but {\it only when or after} kaons condense.

Our approach will be anchored on a multi-skyrmion description of dense hadronic matter~\cite{multifacet}. The merit of our approach is that all the relevant degrees of freedom, i.e., baryons and mesons, in the energy regime concerned, can be treated on the same footing in a unified way  using a single effective Lagrangian that models QCD for both elementary and many-body processes. The approach is quite involved, however, in concepts as well as in numerics, so the results obtained therefrom can be rigorously justified only in the limit of large $N_c$.

The main ingredients of our approach are (1) the Callan-Klebanov (CK) bound-state picture for hyperons formed with (anti-)kaons bound to $SU(2)$ skyrmions~\cite{callan-klebanov} and (2) multiskyrmions put on crystal lattice to simulate dense baryonic matter~\cite{multifacet}.
In the framework so defined, our main result stated above will turn out to be very simple to establish. We will first establish it to ${\cal O}(1)$ in the $N_c$ counting and then discuss what may happen at ${\cal O}(1/N_c)$.
\vskip 0.1cm
{\it Callan-Klebanov Bound State Model for Hyperons:}
In the CK model, (anti-)kaons, taken to be ``heavy," are bound to the $SU(2)$ skyrmion constructed with the Skyrme Lagrangian~\cite{skyrme} consisting of the current algebra term and the four-derivative Skyrme term. The binding is chiefly via the Wess-Zumino term -- related to the Weinberg-Tomozawa term in chiral Lagrangian -- which distinguishes kaon (i.e., $K^+$) and anti-kaon (i.e. $K^-$), pushing the $K^+$ mass up and the $K^-$ mass down.  For the given``heavy-meson" ansatz for the kaon fluctuation, this model does not appear to have flavor $SU(3)$ symmetry. However it turns out that one can recover both chiral $SU(3)\times SU(3)$ symmetry and the (Isgur-Wise) heavy quark  symmetry by tuning the symmetry breaking, i.e., the kaon mass.  An important point to note here that is highly relevant for, and will be exploited in, what follows is that in the limit $m_K\rightarrow 0$, the bound-state description of hyperons smoothly approaches the rigid rotator description~\cite{kaplan-klebanov}. On the other hand as $m_K\rightarrow \infty$,  the heavy-quark symmetry for the s quark emerges~\cite{CND}. We will exploit the former feature in applying the model to kaon fluctuation in dense matter.

The CK model was found to be fairly successful in describing the non-exotic hyperons (with strangeness $S=-1, -2, -3$).  It has the merit to account for the ${\cal O} (N^0)$ strangeness fluctuation that is absent in the rigid rotor model~\cite{klebanov-theta}. That one can single out the strangeness fluctuation is significant since the flavor-independent ${\cal O} (N^0)$ correction, i.e., the Casimir energy, that is difficult to calculate reliably can be eliminated from the consideration. The bound state model can be made to work even better phenomenologically by incorporating vector meson degrees of freedom in the $SU(2)$ sector \`a la hidden local symmetry (HLS)~\cite{scoccola}. Moreover the recent development in applied string theory points to that a more realistic description of baryonic systems could be obtained with the incorporation of the infinite tower of hidden local vector mesons inherent in the gauge-gravity duality~\cite{SS,HRYY}. This is manifested in the striking role that the tower plays in the vector dominance in the meson and baryon form factors~\cite{SS,HRYY}. Now once the vector mesons are present in the $SU(2)$ sector, the homogeneous Wess-Zumino terms inherited from the Chern-Simons action in the 5D  holographic dual action plays an extremely important role in capturing the physics of short-distance strong interactions. There the role of the $\omega$ meson in the baryon structure is found to be crucial~\cite{YLMetal}.

In calculating baryon properties in the skyrmion model, one is resorting to the large $N_c$ expansion of the baryon mass $m=M_1 + M_0 +M_{-1} +\cdots$ where the subscripts stand for the power $n$ in the $N^n_c$ dependence. The first is the leading ${\cal O} (N_c)$ soliton contribution and the third is the ${\cal O}(1/N_c)$ hyperfine splitting obtained by collective-quantization. Both are well defined in the framework with a given Lagrangian. The second term is less straightforward. There are two possible contributions to the ${\cal O}(N_c^0)$ term, $M_0=M_0^0 +M_0^s$. The first is the flavor-independent Casimir energy which is present in both the CK model and the rigid rotor model. The second is the kaonic fluctuation that depends on the strangeness $S$. This term is present in the SK model but absent in the rigid rotor model. It is this term that will play a key role in our discussion that follows.

The Casimir energy is highly model-dependent and hence is difficult to compute reliably. For this reason, it is mostly ignored in the literature. We claim that ignoring this term can be a serious defect in dense matter. Approximate loop calculations using chiral perturbation theory give $M_0\lsim -(1/3) M_1$~\cite{walliser} in matter-free space. In medium, this contribution can be associated with the attraction brought in by a scalar isoscalar field (commonly denoted $\sigma$) that binds nucleons and hence is indispensable in nuclear processes.

In the CK model, the mass formula for hyperon with strangeness $S$ is of the form,
\be
m_{|S|}=M_{sol}+M^0_0+|S|\omega_K +{\cal O}(1/N_c)+\cdots\label{mass}
\ee
where $\omega_K$ is the energy of the (anti)kaon bound -- chiefly by the Wess-Zumino term -- to the soliton. We will be interested in the mass difference between the lowest-lying hyperon -- that we shall denote $Y$ -- and the nucleon. The first two terms of (\ref{mass}) are flavor-singlet and hence are cancelled out in the difference. So taking $|S|=1$ for the lowest-lying hyperon, we obtain the simple result
\be
\Delta m\equiv m_Y-m_N=\omega_K +{\cal O}(1/N_c).
\ee
As will be explained later, the ${\cal O} (1/N_c)$ corrections that lead to the hyperon-multiplet structure are highly model-dependent and cannot be calculated reliably at present. We will therefore limit ourselves to ${\cal O}(N_c^0)$ to which the treatment is robust.
\vskip 0.1cm
{\it The CK Skyrmions on Crystal:}
We are interested in studying kaonic fluctuations in a baryonic medium at high density. For this we put $K^-$s (that we shall simply refer to as kaons -- omitting anti --  for short unless otherwise noted) on crystal lattice following the approach first discussed in \cite{PKR}. The lattice size then defines the matter density. This is the only method we know of for describing in-medium kaons in a unified framework that treats mesons and baryons (both strange and non-strange) in multibaryonic matter with a single Lagrangian with symmetries (i.e., chiral symmetry, scale symmetry) consistent with QCD.  The vector mesons $\rho$, $\omega$ can be incorporated -- naturally and readily -- by hidden local symmetry~\cite{HY:PR}. To capture most, if not all, of the scalar degree of freedom associated with the Casimir effect mentioned above, a dilaton field $\chi$ is introduced via QCD trace anomaly to be {\it treated in mean field} as was suggested for dense matter in \cite{BR91,LR}.

Putting on crystal lattice the HLS Lagrangian implemented with the $\chi$ field  is yet to be worked out. This will yield a more reliable result than available up to date, particularly for the observables sensitive to higher orders in $1/N_c$. Since we are working to ${\cal O}(N_c^0)$, however, we can work with  the simplified Lagrangian of \cite{PKR} where the vector mesons are integrated out. This can be justified for moderate density away from the VM (vector manifestation) density of HLS.
%\footnote{Near the VM fixed point,  the vector meson masses cannot be taken heavy compared with the pion mass, so they cannot be integrated out~\cite{HY:PR}.}

In \cite{PKR}, in close analogy to the CK hyperon, kaon fluctuation $U_K$ in the background of multiskyrmions simulated on FCC crystal $U_0$ was considered by taking the ansatz for the chiral field
\be
U(\vec{x},t)=\sqrt{U_K(\vec{x},t)}U_0(\vec{x})\sqrt{U_K(\vec{x},t)}.\label{ansatz}
\ee
\begin{figure}[here]
\scalebox{0.20}{\includegraphics{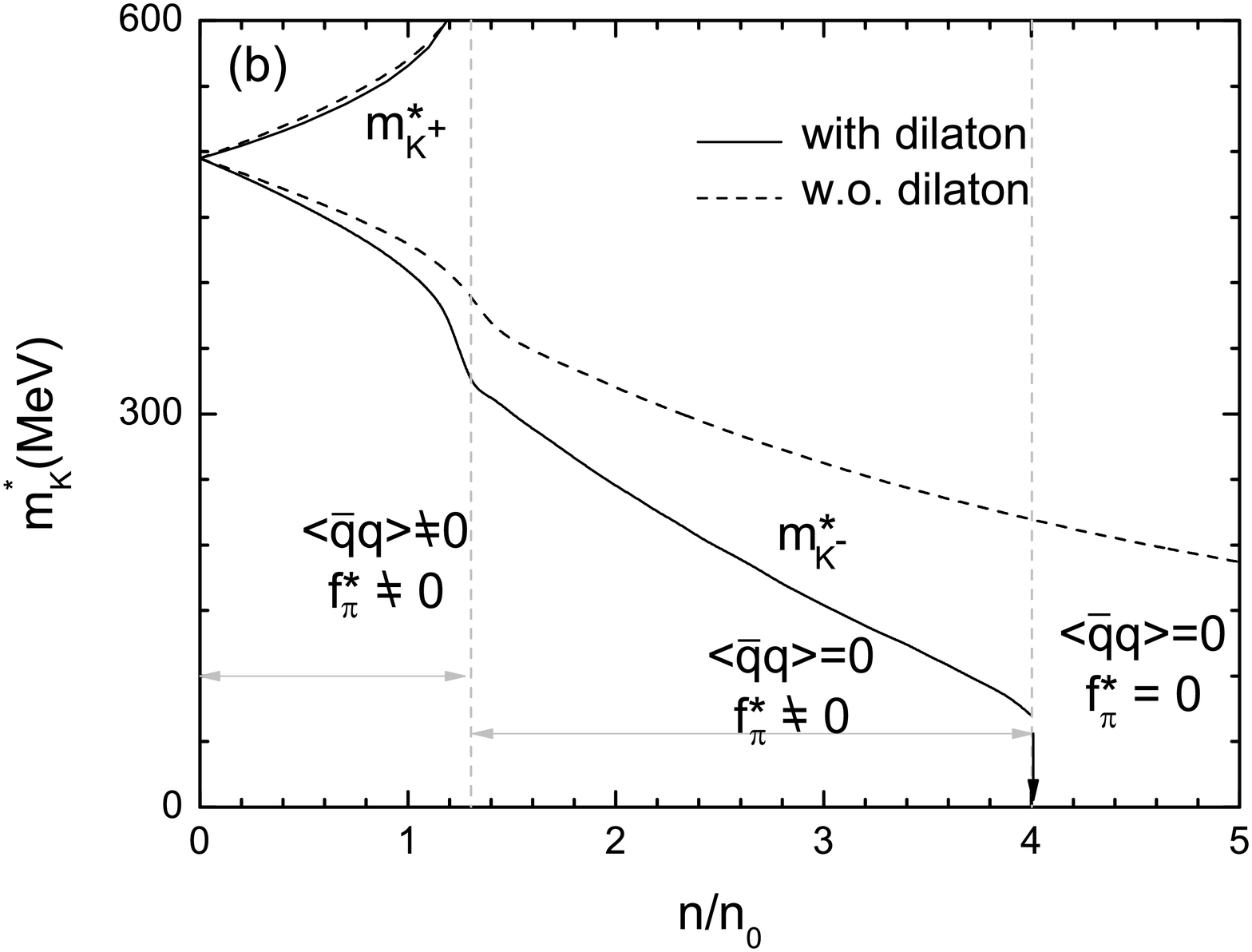}}
\caption{ $m^*_{K^\pm}$ vs. $n/n_0$ in dense skyrmion matter consisting of three phases: (I) skyrmion phase, (II) half-skyrmion phase and
(III) chiral symmetry restored phase. The parameters are fixed
at $\sqrt{2}ef_\pi=780$ MeV and dilaton mass $m_\chi=720$ MeV with $n_{1/2}\sim 1.3n_0$. }
\label{fig}
\end{figure}
There the focus was on the properties of kaons in the given background $U_0$ already calculated in previous studies~\cite{sliding}. A striking qualitative feature found in \cite{sliding} was the skyrmion-half-skyrmion transition that takes place at $n_{1/2}\sim (1.3-2.0)n_0$, a range of density that is compatible with nuclear phenomenology~\cite{dongetal}. While the location of $n_{1/2}$ was found to be insensitive to the dilaton mass, the behavior of the in-medium kaon depended sensitively on the dilaton mass, reflecting the importance of ${\cal O}(N_c^0)$ effects in medium. In Fig.~\ref{fig} is reproduced the scaling of the kaon mass as a function of density for the dilaton mass in the vacuum $m_\chi\approx 720$ MeV~\footnote{This is roughly the vacuum mass for the $\sigma$ that would yield the in-medium mass needed for nuclear matter in mean-field models.}.
 Note that there is an abrupt drop in the $K^-$ mass at $n_{1/2}$, probably connected to the abrupt change in the nuclear tensor forces that affects the symmetry energy in the EoS of compact star matter seen in \cite{dongetal}.

Let us now turn the focus from kaon properties to hyperon properties in medium described as CK skyrmion matter. For this we will ignore the possible back-reaction of the kaons on the background of skyrmions that is suppressed by $1/N_c$ and consider $K^-$'s embedded in an $SU(2)$ skyrmion matter. The ansatz we take is the same as in (\ref{ansatz}). The skyrmion matter with baryon number $B$ is described by the $U_0$ simulated on FCC crystal. As with the CK hyperon for $B=1$, the difference in energy per baryon between the hyperon matter with $|S|=A$ with $K^-$'s bound to the skyrmion matter with baryon number $B=A\gg 1$ and the non-strange skyrmion  matter (with $B=A$ and $|S|=0$) then will be given to ${\cal O} (N_c^0)$ by
\be
E^*_Y -E^*_N=\omega_K^* + {\cal O}(N_c^{-1})\label{inmedium}
\ee
where the asterisk represents medium-dependence. As in the case of $B=1$, the soliton (${\cal O} (N_c)$) and Casimir (${\cal O}(N_c^0)$) energies will cancel out in the difference, leaving only the strangeness-dependent  term, i.e., the in-medium kaon mass, to $O(N_c^0)$. The mass shift occurs as depicted in Fig.~\ref{fig}~\cite{PKR}. The ${\cal O}(N_c^{-1})$  term will capture the hyperfine effect as well as other medium-dependent terms. For this we need to collective-quantize the skyrmion matter with baryon number $A$. The result will be strongly model-dependent such as the degrees of freedom taken into account in the Lagrangian and quantum loop corrections of the Casimir type. At present we do not know how to collective-quantize such multi-skyrmion systems in general. We will mention later what one finds for the simple case where the skyrmion matter is a pure neutron matter. In this case an approximate collective-quantization is feasible~\cite{LPR}. What is of significance is that the treatment given here is robust to the NLO  in $N_c$.
\vskip 0.1cm
{\it Appearance of Strangeness in Compact-Star Matter:}
To see the implication of what we found above in compact-star matter, we recall that according to the scenario of \cite{BTKR}, (s-wave) kaons condense in dense neutron-rich matter when the electron chemical potential $\mu_e$ that increases with increasing baryonic density is equal to or greater than the dropping in-medium kaon mass
\be
\mu_e\geq m_K^*.
\ee
On the other hand, hyperons can appear when
\be
\mu_e\geq E_Y^* -E_N^*.
\ee
From Eq.(\ref{inmedium}), it follows that to NLO in $N_c$, the hyperons will figure {\it at the same density} as the threshold density at which s-wave kaons condense. This is our principal result announced above.

This result could mean either that hyperons and condensed kaons play the same role in compact-star matter or hyperons appear only when kaons condense. For the latter, it would be essential to figure out how hyperons interact in kaon-condensed matter where isospin symmetry is spontaneously broken.  The structure of such strange hadronic matter is an open problem that has not been addressed up to today.
\vskip 0.1cm
{\it Hyperfine Effects:}
Our discussion so far has been limited to ${\cal O} (N_c^0)$. Let us see what one can say about ${\cal O} (1/N_c)$ corrections. These corrections are responsible for the hyperon multiplet structure as well as the splitting between even-parity and odd-parity states\footnote{An important odd-parity hyperon that is often considered for kaon-nuclear physics is $\Lambda (1405)$. This is correctly given by the hyperfine correction.}. Since we do not know how to collective-quantize the skyrmion matter with baryon number $A\gg 1$, we look at the system as a collection of $A$ quasiparticles (i.e., ``quasi-hyperons").  The hyperfine energy for the lowest-lying quasi-hyperon $\Lambda$ of spin $1/2$, isospin $0$ and strangeness $-1$ is~\cite{callan-klebanov,scoccola}
\be
E^*_{-1} (\Lambda)=\frac{3}{8\Omega^*}({c^*}^2-1)\label{massformula}
\ee
where the asterisk stands for in-medium quantity, $\Omega^*$ is the moment of inertia of the skyrmion rotator and  $c^*$ is the coefficient multiplying the effective spin operator of $S=-1$. Taking into account the hyperfine term for the nucleon, we get
\be
E_Y^*-E_N^*=\omega^*_K +\frac{3}{8\Omega^*}({c*}^2-1).\label{med-hyper}
\ee
What this says is that which comes first, hyperons or kaon condensation, will depend on whether ${c*}^2 >1$ or ${c*}^2 <1$. As stressed by Callan and Klebanov, this quantity is highly model-dependent, and cannot be calculated with reliability; one is unable to determine it even for hyperons in the vacuum~\cite{callan-klebanov}. If one uses the mass formula to determine $c$ in matter-free space, it comes out to be $c\sim 0.7$.  If this held in medium, then (\ref{med-hyper}) would imply that hyperons appear {\it before} kaons condense. On the other hand, as mentioned, the CK model smoothly approaches the $SU(3)$ rotor model as $m_K\rightarrow 0$.   In this limit $c^*\rightarrow 1$. In medium, the kaon mass drops, accentuated by the topology change, as density increases, which means that in-medium effects will become more important at ${\cal O} (1/N_c)$ and medium-dependent loop corrections can overwhelm tree contributions. All that can be said at present is that $c^*$ is likely to be close to 1 at high density.
\vskip 0.1cm
{\it Symmetry Energy:}
As done in \cite{LPR} following Klebanov's suggestion~\cite{klebanov}, we can approximately calculate how the symmetry energy is modified in the presence of strangeness by collectively-quantizing  the pure $A$-neutron crystal system (for $A\rightarrow \infty$) with the maximum isospin $I=|I_3|=A/2$. In \cite{LPR}, the symmetry energy of pure neutron matter in the absence of hyperons (or equivalently kaon condensation)  was found to be $E_{sym} =\frac{1}{8{\tilde{\Omega}}}\delta^2$ where $\delta=(N-P)/(N+P)$ and $\tilde{\Omega}$ is the moment of inertia of the unit cell of the crystal lattice. Doing the same calculation with the ansatz (\ref{ansatz}) that includes kaon fluctuation, we obtain
\be
E^{hyp}_{sym}\approx \frac{1}{8\tilde{\Omega}}(1-d^*)\delta^2\label{esym}
\ee
where $d^*$ is what corresponds to $c^*$ in (\ref{med-hyper}) with $U_0$ in (\ref{ansatz}) replaced by pure neutron matter and $\tilde{\Omega}$ is the moment of inertia of the soliton $U_0$. 

A few remarks are in order here.

 Note first that at ${\cal O}(N_c^{-1})$ at which $E_{sym}$ figures, there are terms independent of $\delta$ but dependent on strangeness that could contribute to the hyperfine effects. These terms as well as the $d^*$ cannot be computed accurately. Second, $\tilde{\Omega}$ is precisely what was computed in \cite{LPR} and hence will have the cusp structure at $n_{1/2}$.  However as shown in \cite{dongetal}, the cusp will be smoothed by nuclear correlations . Third, for $d^*>0$, the presence of hyperons will make the symmetry energy softer -- as expected -- and in the limit $m_K^*\rightarrow 0$, we expect $d^*\rightarrow 1$, so the symmetry energy will vanish.

Note also that Eq.(\ref{esym}) will be modified once kaons are condensed, because then isospin is spontaneously broken and there will be a term linear in $\delta$ to the symmetry energy discussed in \cite{LR-flavor}. This means that the description in terms of hyperons -- without condensed kaons -- can no longer be correct.
\vskip 0.1cm
{\it Comments:}
The main conclusion reached in this paper is that to ${\cal O} (N_c^0)$, the appearance of strangeness in compact star matter can be equivalently described either in terms of  hyperons or in terms of  kaons. This is robust to NLO. However, kaon back-reactions will take place at next order in $1/N_c$ as do the hyperfine effects. Without calculating these higher order corrections with confidence, it would be difficult to exclude the possibility of the reversed appearance of the strangeness degrees of freedom.

 As illustrated in Figure~\ref{fig},  the kaon mass in the skyrmion matter drops continuously up to near nuclear matter density which is more or less in accordance with chiral perturbation theory (ChPT) predictions. But the steep drop at $n_{1/2}$ is not seen in ChPT. This drop may have some connection with the Akaishi-Yamazaki ``contraction effect" needed to form dense kaonic nuclei but not present in conventional nuclear interactions~\cite{yamazaki}. It is also interesting to note that with the topology change translated into parameter changes in effective nuclear many-body formulations as discussed in \cite{dongetal}, the dropping of the kaon mass after the density $n_{1/2}$ is close to what one gets for kaon fluctuation in compact stars calculated from the VM fixed point~\cite{VMfluctuation}.

 The topology change at $n_{1/2}$ with the drop of the kaon mass will speed up kaon condensation and soften  the EoS.  On the other hand, in the absence of kaon condensation (or hyperons in our description), it is found to stiffen  the EoS at $n_{1/2}$~\cite{dongetal}. To accommodate both the soft EoS found in heavy-ion experiments at low density and the stiff EoS needed at higher density to account for the recently measured 1.97 $M_\odot$ neutron star, the two mechanisms must intervene in a delicate balancing act. This, we believe, throws open wide the issue of EoS at densities relevant to compact stars.
\vskip 0.1cm
%
%%%%%%%%%%%%%%%%%%%%%%%%%%%%%%%%%%%%%%%%%%%%%%%%%%%%%%%%
{\it Acknowledgments:}
%%%%%%%%%%%%%%%%%%%%%%%%%%%%%%%%%%%%%%%%%%%%%%%%%%%%%%
 The work reported here was partially supported by the WCU project of Korean Ministry of Education, Science and Technology (R33-2008-000-10087-0).
%{ \section*{Appendix: How to get Eq. (\ref{esym})} ``In the moduli-space approximation to  multi-skyrmion dynamics, the Hamiltonian becomes that of a rigid body in space and isospace. So for the minimal energy to the skyrmion, the rotation has the same effect as the isorotation. In other words the combined rotation and isorotation leaves the configuration unchanged." (See P. Irwin, Phys. Rev. {\bf D61} (2000) page 114024-2.)  In this calculation, we are doing the quantization as in CK taking the soliton sector to correspond to pure neutron state. Then the $1/N_c$ contribution to the energy is \be E_{-1}=\frac{(\mathbf{J}_{\rm sol}+d^*\mathbf{J}_K)^2}{2\tilde{\Omega}}.\label{hf} \ee The total angular momentum of the kaon-soliton system is $\mathbf{J}=\mathbf{J}_{\rm sol} +\mathbf{J}_K$ with the subscripts ``sol" and ``K" standing, respectively, for the soliton and the kaon, the total isospin  $|\mathbf{I}|=|\mathbf{I}_{\rm sol}|=|\mathbf{J}_{\rm sol}|$ (kaon carries no isospin in the CK model) and $|\mathbf{J}_K|=|S|/2$ where $|S|$ is the total strangeness. We have \be E_{-1}=\frac{d^* \mathbf{J}^2 +(1-d^*)\mathbf{I}^2 +d^*(d^*-1) \mathbf{J}_K^2}{2\tilde{\Omega}}.\label{hf} \ee From the isospin term follows Eq. (\ref{esym}) proportional to $\delta^2$.  The terms proportional to $\mathbf{J}^2$ and $S^2$ in (\ref{hf}) will go into the hyperfine energy.

\vskip 0.1cm
%%%%%%%%%%%%%%%%%%%%%%%%%%%%%%%%%%%%%%%%%%%%%%%%%%%%%%%%
%%%%%%%%%%%%%%%%%%%%%%%%%%%%%%%%%%%%%%%%%%%%%%%%%%%%%%%%%

\end{document}